# The Astrobiology of the Anthropocene

Jacob Haqq-Misra*, Sanjoy Som, Brendan Mullan, Rafael Loureiro, Edward Schwieterman,
Lauren Seyler, Haritina Mogosanu
*Blue Marble Space Institute of Science*

&

Adam Frank, *University of Rochester*
Eric Wolf, *University of Colorado*
Duncan Forgan, *University of St Andrews*
Charles Cockell, *University of Edinburgh*
Woodruff Sullivan, *University of Washington*

A white paper on "Astrobiology Science Strategy for the Search for Life in the Universe" for the National Academy of Sciences

*Corresponding author
    Address:    Blue Marble Space Institute of Science
                   1001 4th Ave, Suite 3201, Seattle WA 98154
    Email:      jacob@bmsis.org.
    Phone:     206-775-8787

**Introduction: This is the Anthropocene**

Human influence on the biosphere has been evident at least since the development of widespread agriculture over 10,000 years ago, and some stratigraphers have suggested that the activities of modern civilization indicate a geological epoch transition. Materials such as plastics, concrete, and other "technofossils" will continue to join the products of fossil fuel combustion as a uniquely anthropogenic contribution to the sedimentary record, while geochemical residues from pesticides and fertilizers will also remain in the rock record ([Waters et al. 2016](#)). Fallout from nuclear tests likewise constitute a detectable signature, which has led to the suggestion of demarcating the Holocene-Anthropocene boundary at the time of the world's first nuclear bomb explosion in 1945 ([Zalasiewicz et al. 2015](#)).

Future changes in the Earth system could also leave stratigraphic signatures. Some modeling studies have suggested that anthropogenic climate change could delay or even halt the ice-age cycle ([Herrero et al. 2014](#); [Haqq-Misra 2014](#)). Proposals to counteract climate change through intentional geoengineering include the idea of actively promoting the growth of thick ice sheets in order to alter Earth's energy balance ([Haqq-Misra 2015](#); [Desch et al. 2017](#)) and injecting sulfate particles into the stratosphere to increase cloud albedo ([Vaughn & Lenton 2011](#); [Moreno-Cruz & Keith 2013](#)), both of which could contribute to changes in stratigraphy. Social instability could also leave a stratigraphic signature, such as the catastrophic consequences following a global nuclear winter ([Robock et al. 2007](#)).

Our perspective as a civilization living within the epoch transition from the Holocene to the Anthropocene allows us to contemplate the emergence of technological civilization in the context of planetary-scale processes ([Grinspoon 2016](#)). The Anthropocene may even represent a predictable planetary transition in general, to the extent that any energy-intensive species should drive changes in its biosphere ([Frank et al. 2017](#)). Examining the Anthropocene epoch through the lens of astrobiology can help to understand the future evolution of life on our planet and the possible evolution of technological, energy-intensive life elsewhere in the universe.

**Climate Change as a Planetary Process**

Climate change is one of the most salient science and political issues of our time. From an astrobiological perspective, drastic climate changes such as the Great Oxygenation Event (or Oxygen Catastrophe) at the beginning of the Proterozoic, the Neoproterozoic Snowball Earth episodes, or the Paleocene-Eocene Thermal Maximum ([McInerney & Wing 2011](#)) have led to major shifts in the dominant forms of life on Earth. The Permian-Triassic extinction event, also known as the "Great Dying," may have been caused in part by the production of methane by the archaea *Methanosarcina* ([Rothman et al. 2014](#)). A methanogen-dominated biosphere may have also generated a protective haze layer during the Archean to maintain habitable conditions ([Arney et al. 2017](#)). Such events illustrate the ability of life to act as a transformative process on a planet, shaping the conditions that will accommodate future lifeforms. Yet Earth has remained continuously inhabited for almost 4 billion years while going through a wide range of immense environmental and atmospheric changes. Methods for inferring properties such as the air density (*e.g.*, [Kavanagh & Goldblatt 2015](#); [Som et al. 2016](#)), temperature (*e.g.* [Valley et al. 2002](#)), and redox state (*e.g.* [Catling & Claire 2005](#)) of Earth's atmosphere through geologic time give snapshots of known inhabited worlds different from modern Earth. The study of Earth's habitability through geologic time provides a basis for understanding how the Earth system will respond to anthropogenic contributions in the future.

Comparative planetology provides another route for understanding climate processes in a broader physical context. Study of the atmospheres of Mars, Venus, and Titan, past and present, provide important observable examples of how planets undergo long-term climate evolution. Theoretical studies of planetary habitability climate tend to push the climate models into non-Earth regimes that are relevant for exoplanet characterization, such as synchronous rotation, extremely high carbon dioxide, and other exotic orbital configurations. The concept of a runaway greenhouse can explain the history of Venus ([Way et al. 2016](#)) and delineate the inner edge of the habitable zone ([Kopparapu et al. 2013](#)), which has also inspired investigation of the threshold at which anthropogenic activity could induce a runaway greenhouse on Earth today ([Goldblatt et al. 2013](#); [Ramirez et al. 2014](#); [Popp et al. 2016](#)). Accurate representation of clouds is important for exoplanet climate models, which has also been identified by the Intergovernmental Panel on Climate Change (IPCC) as a critical area of improvement needed for Earth models. The exoplanet climate modeling community has made significant progress over the past five years toward improving our understanding of large-scale cloud processes for synchronously rotating planets ([Yang et al. 2013](#)), runaway greenhouse thresholds ([Leconte et al. 2013](#)), and other habitability constraints (*e.g.*, [Fujii et al. 2017](#); [Kopparapu et al. 2017](#); [Turbet et al. 2017](#)).

Exoplanet atmospheric characterization is a related emerging area of interest that provides additional data for understanding how atmospheres evolve, which includes recent discoveries like Proxima Centauri b, the TRAPPIST-1 system, LHS1040b, and Ross 128b. Comparative modeling studies within the exoplanet science community could also benefit from interdisciplinary collaboration with the Earth climate community, especially to support mutual model development goals.

Future projections of climate change on Earth show a range of likely outcomes, which depends in part upon humanity's response to engage in mitigation, adaptation, and (perhaps) geoengineering. Geoengineering research is not within the scope of the astrobiology program, but interdisciplinary collaboration with geoengineering groups could also lead to progress on both present-day climate and exoplanet habitability problems. One possible link is understanding the role of geoengineering in the long-term future of Earth ([Goldblatt & Watson 2012](#)), which could help to predict potential remote signatures of geoengineering on exoplanets.

**The Limits and Lifetime of Human Civilization**

Beyond the present-day climate problem, any growing technological civilization living on a finite planet will face limits and consequences to growth, while enduring self-induced or extant threats that compound with time. This realization prompted Thomas Malthus' warnings in 1789 about the limits of human population growth and the effects on the environment and agricultural systems, which suggested that a civilizational collapse was possible. Subsequent analyses in the following centuries adjusted the threshold of this collapse based upon the ability of technology to raise a planet's carrying capacity.

A modern analysis of this problem with recent population and agricultural data finds that growth in the food supply should continue to outpace current trends in population in the foreseeable future (Mullan & Haqq-Misra 2018). However, this analysis also finds that, even if greenhouse gas emissions are mitigated, growth in human civilization's energy use will thermodynamically continue to raise Earth's equilibrium temperature. If current energy consumption trends continue, then ecologically catastrophic warming beyond the heat stress tolerance of animals (Sherwood & Huber 2010) may occur by ~2200-2400, independent of the predicted slowdown in population growth by 2100 (Raftery 2012).

The limits imposed by thermodynamics on a growing civilization suggest that such effects could be a universal feature of planets that have undergone an Anthropocene-like transition (Frank et al. 2017). The predicted fractional change in temperature ($\Delta T/T \sim 2\%$) at this thermodynamic limit corresponds to a world power use of $O(10^{16})$ W (compared to the $O(10^{13})$ W for today), or about 7% of the incoming solar radiation (Mullan & Haqq-Misra 2018). Further work is needed to determine whether this is a hard limit to energy consumption, or whether this limit depends on second order climate effects, ecosystem stability, atmospheric composition, orbital distance, planetary radius, or other properties of a civilization's host planet.

The existence of thermodynamic growth limits for human civilization also suggests possible explanations for the Fermi paradox and strategies in the search for extraterrestrial intelligence (SETI). Calculations of the mean lifetime of energy-intensive civilizations help place general planetary habitability models into the specific context of civilizational habitability, which implies that sustainability limits should apply in general to civilizations everywhere (Frank & Sullivan 2014). The $O(10^{16})$ W energy limit coincidentally corresponds to a "Type-I" civilization according to the Kardashev scale, which may point toward a fundamental limit of the observational imprint of a developing civilization.

Other risk factors could also reduce the longevity of a technological civilization, which are often collectively referred to as global catastrophic risks (Bostrom & Cirkovic 2008) or existential risks (Bostrom 2013). Such possibilities include nuclear winter, pandemic, and asteroid impacts, as well as projected catastrophic failures of future technologies, such as artificial intelligence. Collaborations with research communities that study global catastrophic and existential risks would help to develop quantitative constraints on the expected mean lifetime of energy-intensive civilizations. Such constraints would improve policy decisions for the future of human civilization and also guide SETI toward targets most likely to host extant civilizations.

**Searching for Other Civilizations**

Civilization and technology emerged once from the planetary processes on Earth, which provides an example of what to look for elsewhere. This does not necessarily imply that other inhabited planets will follow the same trajectory as life on Earth. Instead, SETI tends to operate with the working hypothesis that anything that happened here on Earth, or that is possible to happen in the future, remains a plausible option for guiding the search for other civilizations.

SETI has so far managed to continue its efforts by appealing to the private sector for funding, such as the Breakthrough Listen initiative ([Enriquez et al. 2017](#)) as well as nightly surveys by the SETI Institute using the Allen Telescope Array ([Harp et al. 2016](#)). SETI represents an important objective of astrobiology, as progress in identifying and characterizing exoplanets also allows SETI to select better targets. This is an area for continued collaboration, in order to allow the observational and theoretical habitability studies from within astrobiology to also benefit SETI research ([Frank & Sullivan 2016](#)).

Spectral signatures provide one way to characterize a planet's atmosphere, with a sizable astrobiology literature on possible atmospheric biosignatures that could indicate the presence of surface life (*e.g.*, [Schwieterman et al. 2018](#)). Spectral technosignatures are a particular spectral signature that would indicate the presence of a technological civilization on the planet ([Schneider et al. 2010](#); [Stevens et al. 2016](#)). Examining the effects of human civilization on Earth's climate, both today and in likely future trajectories, can help to identify plausible technosignatures that might be observed with the next generation of space telescopes.

The terraforming of otherwise uninhabitable planets within a planetary system is one example of a possible technosignature, where powerful artificial greenhouse gases may be deployed to warm a planet outside the formal habitable zone ([Fogg 2010](#)). Such planets may be identified from the spectral features of greenhouse gases such as perfluorocarbons (PFCs), which are not known to otherwise occur in high abundances. Spectral technosignatures would produce the most observable features in the infrared portions of the electromagnetic spectrum, specifically in the thermal infrared window region at 8-12 μm for greenhouse gases. Conceptual studies of space telescopes capable of imaging terrestrial planets in the mid-infrared were previously studied, such as NASA's Terrestrial Planet Finder Infrared (TPF-I) space-based interferometer design concept ([Beichman et al. 2006](#)), and ESA's Darwin concept ([Cockell et al. 2009](#)), although neither is currently under consideration by either agency. The Origins Space Telescope (OST) concept is currently under study, which could resolve terrestrial planet features in the 8-12 μm range ([Cooray et al. 2017](#)).

The search for megastructures, such as Dyson swarms or other artifacts of extraterrestrial engineering, complements existing spectral surveys. The observation of anomalous absorption in the KIC 8462852 system (also known as "Boyajian's Star") prompted speculation on the possibility of detecting megastructures through transit photometry ([Wright & Sigurdsson 2016](#); [Gaidos 2017](#)). Astrobiologists may therefore inevitably find themselves part of this discussion, particularly if future missions detect other unusual transit or spectral features.

**Sustaining the Overview Effect**

The "overview effect" is a feeling described by astronauts as a cognitive shift in awareness that comes from viewing the Earth from space. A common expression of the overview effect is a profound understanding of the interconnection of all life and a renewed sense of responsibility for taking care of our planet ([White 2014](#)). Even for those not fortunate enough to experience the view firsthand, the overview effect can still be expressed and felt by standing on Earth's surface. This was perhaps most poignant with the release of the iconic Earth-from-space images from Apollo 8 and Apollo 11 that continue to endure in popularity today ([Chaikin 2007](#)).

Images convey emotions that words cannot, and modern space missions all have cameras for this reason. Carl Sagan's "[Pale Blue Dot](#)" image of Earth taken from Voyager 1 inspired other initiatives, such as the "[Pale Blue Orb](#)" image of Earth taken from Cassini. Similar impressions of awe and wonder occur when viewing the [Hubble Deep Field](#) images. Improvements in data bandwidth technologies have also led to new Earth-observing platforms. For example, the [Japanese Himawari 8](#) and [American DISCOVR](#) weather satellites are uniquely positioned to observe the whole terrestrial disk and operate websites for the public to view real-time Earth images. The [International Space Station](#) also broadcasts a live image stream of Earth from space.

The Large Ultraviolet Optical and InfraRed (LUVOIR), Habitable Exoplanet Imaging Mission (HabEx), and OST are three space telescope concepts currently under study by NASA that would provide unprecedented advances in the study of exoplanets and extragalactic astronomy ([Dalcanton et al. 2015](#); [Bolcar et al. 2017](#); [Mennesson et al. 2016](#); [Crill & Siegler 2017](#)), both of which are areas with notable public interest. If any of these next generation observatories is deployed, then consideration should be given to observations of images that resonate with the public, much as the Hubble Deep Field images provided broad appeal beyond immediate science goals.

**Conclusion and Recommendations**

The study of the anthropocene as a geological epoch, and its implication for the future of civilizations, is an emerging transdisciplinary field in which astrobiology can play a leading role. We recommend two approaches toward making significant progress in this area:

- The NASA Astrobiology Institute should establish a [Focus Group](#) on the "astrobiology of the Anthropocene." This focus group would develop a subcommunity of scholars interested in studying Earth's future, drawing from within the astrobiology community as well as drawing upon other experts from the climate change, geoengineering, SETI, security, education, and risk communities.

- NASA should maintain the development of missions such as LUVIOR, HabEx, and OST, which will provide the best opportunity in the coming decades to observe terrestrial biosignatures. Future decadal surveys should consider mission concepts similar to TPF-I and Darwin, or even a lunar observatory, in order to characterize biosignatures and possible technosignatures in the thermal infrared region.

**References**

Arney, G. N., Meadows, V. S., Domagal-Goldman, S. D., et al. (2017). Pale orange dots: the impact of organic haze on the habitability and detectability of Earthlike exoplanets. The Astrophysical Journal, 836(1), 49.

Beichman, C., Lawson, P., Lay, O., Ahmed, A., Unwin, S., & Johnston, K. (2006). Status of the terrestrial planet finder interferometer (TPF-I).

Bolcar, M. R., Aloezos, S., Bly, V. T., et al. (2017, September). The large uv/optical/infrared surveyor (luvoir): Decadal mission concept design update. In UV/Optical/IR Space Telescopes and Instruments: Innovative Technologies and Concepts VIII (Vol. 10398, p. 1039809). International Society for Optics and Photonics.

Bostrom, N., & Cirkovic, M. M. (Eds.). (2011). Global catastrophic risks. Oxford University Press.

Bostrom, N. (2013). Existential risk prevention as global priority. Global Policy, 4(1), 15-31.

Catling, D. C., & Claire, M. W. (2005). How Earth's atmosphere evolved to an oxic state: a status report. Earth and Planetary Science Letters, 237(1), 1-20.

Chaikin, A. (2007). Live from the Moon: the societal impact of Apollo. Societal Impact of Spaceflight, 56.

Cockell, C. S., Léger, A., Fridlund, M., et al. (2009). Darwin—a mission to detect and search for life on extrasolar planets.

Cooray, A. R., & Origins Space Telescope Study Team. (2017, January). Origins Space Telescope. In American Astronomical Society Meeting Abstracts (Vol. 229).

Crill, B. P., & Siegler, N. (2017). Space technology for directly imaging and characterizing exo-Earths. In UV/Optical/IR Space Telescopes and Instruments: Innovative Technologies and Concepts VIII (Vol. 10398, p. 103980H). International Society for Optics and Photonics.

Dalcanton, J., Seager, S., Aigrain, S., et al. (2015). From cosmic birth to living earths: the future of UVOIR space astronomy. arXiv preprint arXiv:1507.04779.

Desch, S. J., Smith, N., Groppi, C., Vargas, P., Jackson, R., Kalyaan, A., ... & Spacek, A. (2017). Arctic ice management. Earth's Future, 5(1), 107-127.

Enriquez, J. E., Siemion, A., Foster, G., et al. (2017). The Breakthrough Listen Search for Intelligent Life: 1.1–1.9 GHz Observations of 692 Nearby Stars. The Astrophysical Journal, 849(2), 104.

Fogg, M. J. (2011). Terraforming Mars: A Review of Concepts. In Engineering Earth (pp. 2217-2225). Springer Netherlands.

Frank, A., & Sullivan, W. (2014). Sustainability and the astrobiological perspective: Framing human futures in a planetary context. Anthropocene, 5, 32-41.

Frank, A., & Sullivan III, W. T. (2016). A new empirical constraint on the prevalence of technological species in the universe. Astrobiology, 16(5), 359-362.

Frank, A., Kleidon, A., & Alberti, M. (2017). Earth as a Hybrid Planet: The Anthropocene in an Evolutionary Astrobiological Context. Anthropocene, 19, 13-21.

Fujii, Y., Del Genio, A. D., & Amundsen, D. S. (2017). NIR-Driven Moist Upper Atmospheres of Synchronously Rotating Temperate Terrestrial Exoplanets. arXiv preprint arXiv:1704.05878.

Gaidos, E. (2017). Transit Detection of a "Starshade" at the Inner Lagrange Point of an Exoplanet. Monthly Notices of the Royal Astronomical Society, 469(4), 4455-4464.

Goldblatt, C., & Watson, A. J. (2012). The runaway greenhouse: implications for future climate change, geoengineering and planetary atmospheres. Phil. Trans. R. Soc. A, 370(1974), 4197-4216.

Goldblatt, C., Robinson, T. D., Zahnle, K. J., & Crisp, D. (2013). Low simulated radiation limit for runaway greenhouse climates. Nature Geoscience, 6(8), 661-667.

Grinspoon, D. (2016). Earth in Human Hands: Shaping Our Planet's Future. Hachette UK.

Haqq-Misra, J. (2014). Damping of glacial-interglacial cycles from anthropogenic forcing. Journal of Advances in Modeling Earth Systems, 6(3), 950-955.

Haqq-Misra, J. (2015). Should we geoengineer larger ice caps? Futures, 72, 80-85.

Herrero, C., García-Olivares, A., & Pelegrí, J. L. (2014). Impact of anthropogenic $CO_2$ on the next glacial cycle. Climatic change, 122(1-2), 283-298.

Kavanagh, L., & Goldblatt, C. (2015). Using raindrops to constrain past atmospheric density. Earth and Planetary Science Letters, 413, 51-58.

Kopparapu, R. K., Ramirez, R., Kasting, J. F., et al. (2013). Habitable zones around main-sequence stars: new estimates. The Astrophysical Journal, 765(2), 131.

Kopparapu, R. K., Wolf, E. T. Arney, G., et al. (2017). Habitable most atmospheres on terrestrial planets near the inner edge of the habitable zone around M dwarfs. The Astrophysical Journal, 845, 5.

Leconte, J., Forget, F., Charnay, B., Wordsworth, R., & Pottier, A. (2013). Increased insolation threshold for runaway greenhouse processes on Earth-like planets. Nature, 504(7479), 268-271.

McInerney, F. A., & Wing, S. L. (2011). The Paleocene-Eocene Thermal Maximum: a perturbation of carbon cycle, climate, and biosphere with implications for the future. Annual Review of Earth and Planetary Sciences, 39, 489-516.

Mennesson, B., Gaudi, S., Seager, S., et al (2016). The Habitable Exoplanet (HabEx) Imaging Mission: preliminary science drivers and technical requirements. In Space Telescopes and Instrumentation 2016: Optical, Infrared, and Millimeter Wave (Vol. 9904, p. 99040L). International Society for Optics and Photonics.

Moreno-Cruz, J. B., & Keith, D. W. (2013). Climate policy under uncertainty: a case for solar geoengineering. Climatic Change, 121(3), 431-444.

Mullan, B., & Haqq-Misra, J. (2018). Population growth, doomsday, and the implications for the search for extraterrestrial intelligence. Futures, in review, preprint at http://goo.gl/E5V6zA

Popp, M., Schmidt, H., & Marotzke, J. (2016). Transition to a Moist Greenhouse with $CO_2$ and solar forcing. Nature communications, 7.

Raftery, A. E., Li, N., Ševčíková, H., Gerland, P., & Heilig, G. K. (2012). Bayesian probabilistic population projections for all countries. Proceedings of the National Academy of Sciences, 109(35), 13915-13921.

Ramirez, R. M., Kopparapu, R. K., Lindner, V., & Kasting, J. F. (2014). Can increased atmospheric CO2 levels trigger a runaway greenhouse?. Astrobiology, 14(8), 714-731.

Robock, A., Oman, L., & Stenchikov, G. L. (2007). Nuclear winter revisited with a modern climate model and current nuclear arsenals: Still catastrophic consequences. Journal of Geophysical Research: Atmospheres, 112(D13).

Rothman, D. H., Fournier, G. P., French, et al. (2014). Methanogenic burst in the end-Permian carbon cycle. Proceedings of the National Academy of Sciences, 111(15), 5462-5467.

Schneider, J., Léger, A., Fridlund, M., et al. (2010). The far future of exoplanet direct characterization. Astrobiology, 10(1), 121-126.

Schwieterman, E. W., Kiang, N. Y., Parenteau, M. N., et al. (2017). Exoplanet Biosignatures: A Review of Remotely Detectable Signs of Life. arXiv preprint arXiv:1705.05791.

Sherwood, S. C., & Huber, M. (2010). An adaptability limit to climate change due to heat stress. Proceedings of the National Academy of Sciences, 107(21), 9552-9555.

Som, S. M., Buick, R., Hagadorn, J. W., et al. (2016). Earth's air pressure 2.7 billion years ago constrained to less than half of modern levels. Nature Geoscience, 9(6), 448-451.

Stevens, A., Forgan, D., & James, J. O. M. (2016). Observational signatures of self-destructive civilizations. International Journal of Astrobiology, 15(4), 333-344.

Turbet, M., Bolmont, E., Leconte, J., et al. (2017). Modelling climate diversity, tidal dynamics and the fate of volatiles on TRAPPIST-1 planets. arXiv preprint arXiv:1707.06927.

Valley, J. W., Peck, W. H., King, E. M., & Wilde, S. A. (2002). A cool early Earth. Geology, 30(4), 351-354.

Vaughan, N. E., & Lenton, T. M. (2011). A review of climate geoengineering proposals. Climatic change, 109(3-4), 745-790.

Waters, C. N., Zalasiewicz, J., Summerhayes, C., et al. (2016). The Anthropocene is functionally and stratigraphically distinct from the Holocene. Science, 351(6269), aad2622.

Way, M. J., Del Genio, A. D., Kiang, N. Y., et al. (2016). Was Venus the first habitable world of our solar system?. Geophysical Research Letters, 43(16), 8376-8383.

White, F. (1998). The overview effect: Space exploration and human evolution. AIAA.

Wright, J. T., & Sigurdsson, S. (2016). Families of Plausible Solutions to the Puzzle of Boyajian's Star. The Astrophysical Journal Letters, 829(1), L3.

Yang, J., Cowan, N. B., & Abbot, D. S. (2013). Stabilizing cloud feedback dramatically expands the habitable zone of tidally locked planets. The Astrophysical Journal Letters, 771(2), L45.

Zalasiewicz, J., Waters, C. N., Williams, M., et al. (2015). When did the Anthropocene begin? A mid-twentieth century boundary level is stratigraphically optimal. Quaternary International, 383, 196-203.